# Computational-Experimental Investigation of a Fission Thermal Probe in TREAT

**Nima Fathi**[*]
University of New Mexico
P.O. Box 4117
87196, Albuquerque, NM, USA
nfathi@unm.edu

Patrick McDaniel [1], Nicolas E. Woolstenhulme [2], Cassiano de Oliveira [1], Lance Hone [2] and Joshua Daw [2]

[1] *Department of Nuclear Engineering, MSC01 1120*
*1 University of New MexicoAlbuquerque, NM, USA*
[2] *Idaho National Laboratory, P.O. Box 1625, MS 6188, Idaho Falls, ID, USA*

mcdanielpk@gmail.com, nicolas.woolstenhulme@inl.gov, cassiano@unm.edu,
lance.hone@inl.gov, joshua.daw@inl.gov

**ABSTRACT**

The development of nuclear fuels requires unsteady/transient testing for design process and qualification under postulated accident conditions. Breach, rupture, fracture, melting, and other fuel failure modes may occur during the use of nuclear fuels or when they are exposed to extreme overpower conditions. Therefore, it is crucial for the fuel to retain reasonable structural integrity and coolable geometry. The experimental facility of Transient Reactor Test (TREAT) was designed and constructed in November 1958 to conduct transient testing of fuels and structural materials. The magnitude of nuclear heating is one of the most important key parameters in test design, analysis, and data interpretation in TREAT. Some steady-state tests are able to measure heating directly via enthalpy rise of coolant in thermally-isolated loops, but an enormous number of important tests must rely upon nuclear modeling to correlate core operating parameters to specimen nuclear heating. Uncertainties of these models and their inputs can prevent experimenters from achieving desired conditions and hamper advanced models from describing the phenomena to the level of confidence needed. The situation can be even more difficult for unsteady/transient tests where nuclear heating is intentionally varied over short time scales. This research develops a novel nuclear heating sensor technology which directly measures the parameters of interest using spatially-resolved real-time thermometry of fissionable instrument materials, demonstrate the instruments' use in TREAT, and perform data comparisons to nuclear models to facilitate an advanced understanding of transient power coupling. Here we present our results of the designed probe and its testing outcome.

---

[*] Presenting/Corresponding Author

# 1      INTRODUCTION

The air-cooled experimental facility of TREAT which was constructed in 1958 is a thermal system to generate a large transient thermal flux in test elements while the core of the reactor is not overheated. This test reactor is composed of graphite blocks encapsulated in canisters of sheet metal zirconium alloy. The uniform dilute concentration of uranium oxide in the fuel blocks provides this testing system with rapidly distributed transient nuclear heating into the graphitic heat-sink/moderator which causes a neutron energy spectral shift with strong negative temperature feedback for safe self-limiting power excursions. A perspective view of TREAT is shown in Figure 1[1-4].

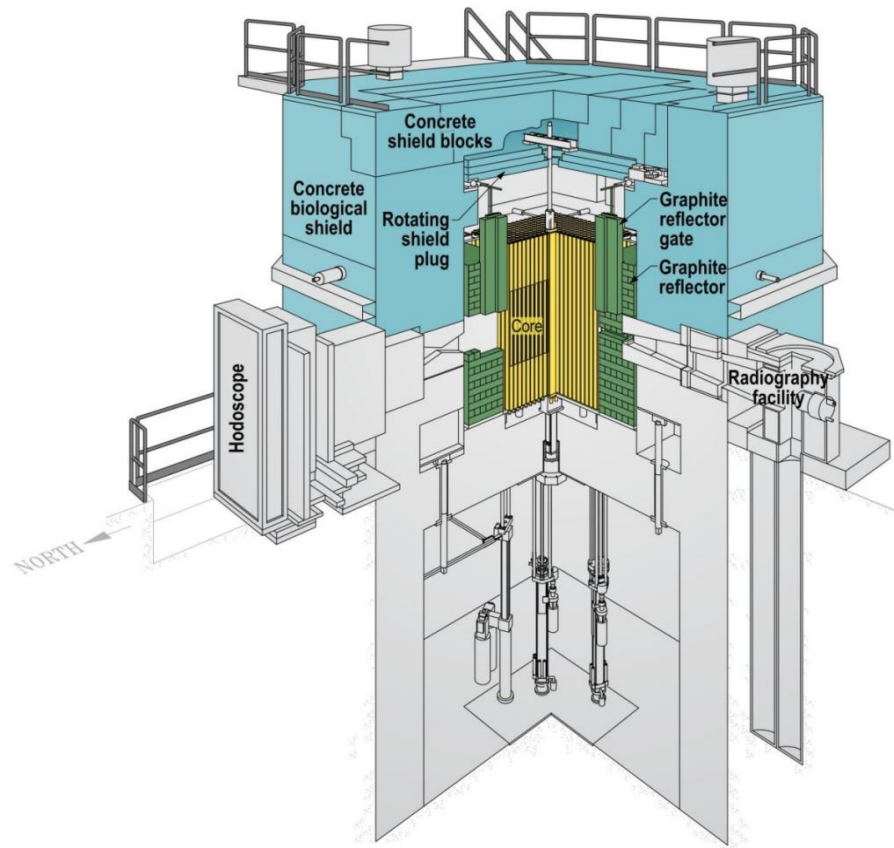

Figure 1: Perspective view of TREAT [2]

The TREAT facility is a very flexible and useful experimental tool allowing its fuel elements and reflectors to be rearranged in many ways.  The central core and reflector area of the reactor consists of a fuel support plat that has a square matrix of 19X19 locations that can accept 4 in by 4 in fuel, control rod, or reflector elements.  This is surrounded by a permanent graphite reflector and then a concrete biological shield. Every location external to the biological shield is accessible during steady state operation. The only restricted area is the control rod drive chamber below the core.  A plan view of the reactor is presented in Figure 2. With all control rods removed, it had a positive excess reactivity of ~60 inhours or a $\Delta k_{excess}$ of 0.001566. All of our analysis is directed at this particular core arrangement as its performance is very well described in ANL-6173[5].

508.3

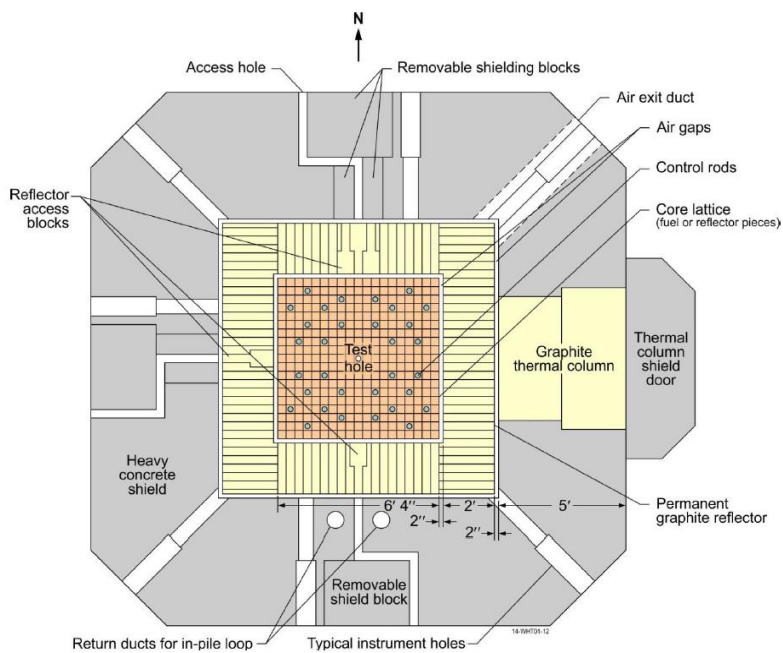

Figure 2: Plan view of the TREAT core [2]

## 2 OBJECTIVE AND CONCEPTUAL DESIGN

While classical dosimeter-type measurements (e.g. radiochemistry and activity counting for flux/fission wires/foils) can be applied to measure neutron spectra/flux and fission rate, they do not directly measure gamma heating, or the amount of fission heating deposited in the dosimeter. Additionally, considerations for decay time, shipment, dose rate, dosimeter quantity/size, and isotopes of interest often compete so that the dosimeter-type measurements are not ideally refined, accurate, or timely. Lastly, and especially relevant for transient experiments, dosimeter-type measurements cannot describe time-dependent nuclear heating. In transient reactors, where spectral and spatial shifts in neutron flux are considerable, dosimeters can be used to account for the time-averaged effect of transient physics. Development and use of the sensor technology addressed in this research elucidate these issues to enable more-accurate experiment design and fuel performance analysis. The selected design will be a novel application of an ultrasonic thermometer where the in-pile sensing element is a high temperature uranium alloy wire insulated for near-adiabatic response in transient irradiations. In this case the sensing element would be a uranium-alloy wire directly heated by the reactor (fission/gamma heat) paired with a non-fissile wire of the same density to separately measure gamma heating. This approach is like activation measurements on uranium-alloy fission wires routinely used in TREAT calibration tests to operational events, especially unplanned and safety events. A drawing of a typical ultrasonic thermometer is presented in Figure 3.

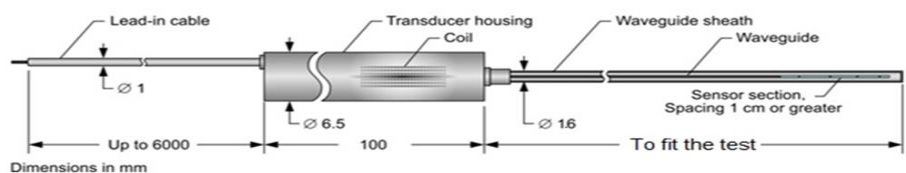

Figure 3: Schematic representation of the probe

## 3  MODELING APPROACH

In order to understand the environment and the performance of this probe, a point kinetics model was developed for the TREAT reactor aimed at providing the heating source term for the probe. The neutronic model is a point one but it contains the six delay families of delayed neutrons and an adiabatic core. The emphasis in this model is the heat deposition and transfer from the probe. The model chosen is a one-dimensional model at a given height in the reactor. All heat transfer is in the radial direction. The model consists of the probe with an outer diameter of 0.0625 in (0.0794 cm radius), and a heat deposition rate of 0.9 watts per megawatt in the TREAT core. This is surrounded by a static air gap (outer radius 0.1216 cm) followed by an MgO insulator layer (outer radius 0.1981 cm) and then a stainless-steel sheath (outer radius 0.2781 cm). The sheath is surrounded by another air layer (outer radius 0.3175 cm) followed by the graphite core of the TREAT reactor. The thermal conductivity of air at the room temperature is $0.0223 \, W/mK$. Given this model, consider what the temperature traces will look like for 4 typical TREAT super-critical pulses. The MgO was installed on the inside of the sheath to inhibit heat transfer out of the probe. However approximately every 4.25 inches (10.795 cm) an MgO spacer is installed to hold the probe in the center of the sheath. Figure 4 represents the thermophysical properties (TPRC) MgO used in our modeling approach. The room temperature thermal conductivity for the MgO is approximately $34 \, W/mK$. Thus, the MgO spacers can contribute to the heat transfer from the probe to the MgO insulator and sheath. The MgO spacers make up approximately 2% of the length of the probe. So, if one calculates an average thermal conductivity, it will include 2% of MgO using a linear approach for calculation of effective material properties. This linear procedure is a very crude way to consider the 2-dimensional heat transfer that occurs. However, the probe itself integrates over its length when it is pulsed, so this is a reasonable first approximation to the total heat transfer. Besides the point kinetics modeling, we have also applied finite element method, and validated our modeling and simulation results against experimental data.

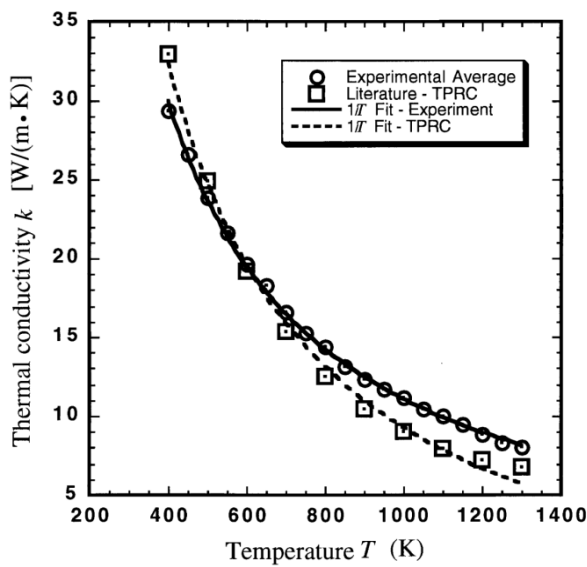

Figure 4: Thermal conductivity of MgO [6,7]

## 4  RESULTS AND DISCUSSION

Given the developed point kinetics model, consider what the temperature traces will look like for 4 typical TREAT super-critical pulses. Consider reactivity inputs of 6.25$, 4.25$, 2.25$

and 1.25$. For a beta-effective of 0.0072 these correspond to 0.0450, 0.0306, 0.0162, and 0.0090 Δk/k. The results for the probe, sheath, and core temperature histories for the 6.25$ pulse are given in Figure 5.

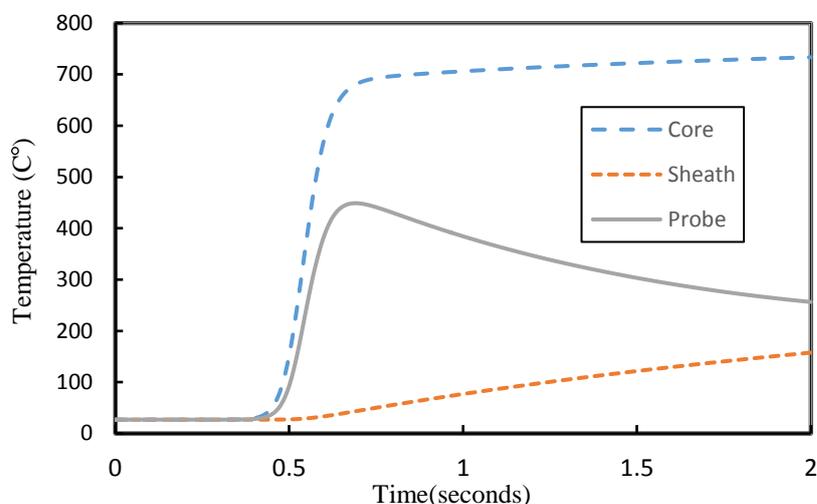

Figure 5: Temperature histories for the probe, sheath and TREAT core during a 6.25$ pulse

Several observations are worth pointing out. First the core temperature exceeds that of the sheath or the probe. This is based on a pure adiabatic pulse and is the average core temperature. At the centre of the core the temperature rise is probably greater. The effect of this higher temperature will need to be explored in the future. Second the temperature measured by the probe rises and then decays. If the electronic pulser that measures probe expansion is operated throughout the transient, it is likely that several pulses will record near the peak given a 10-20 millisecond spacing between pulses. The temperature traces for the 4.25$ pulse are plotted in Figure 6.

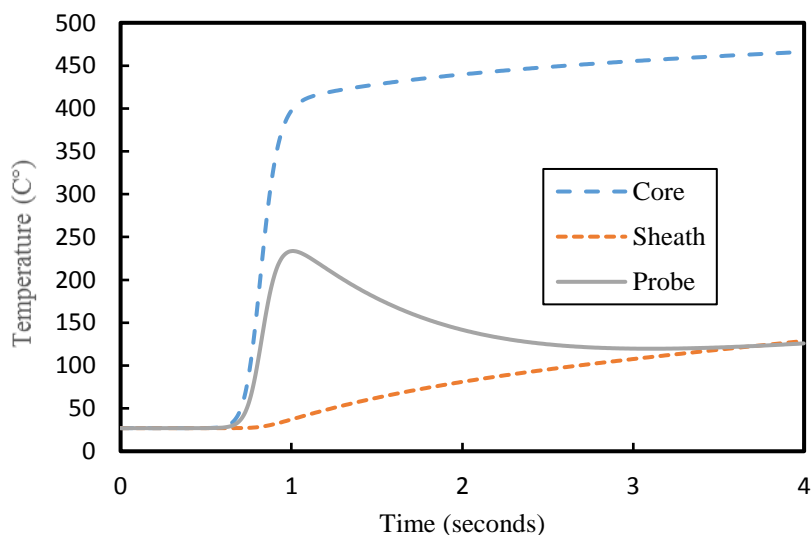

Figure 6: Temperature histories for the probe, sheath and TREAT core during a 4.25$ pulse

The time histories for this case look very similar except the probe temperature history has merged with the sheath temperature history by about the 3.5 second point. The sheath is heated both by the probe and by the reactor core. The time histories for the 2.25$ pulse are given in Figure 4.

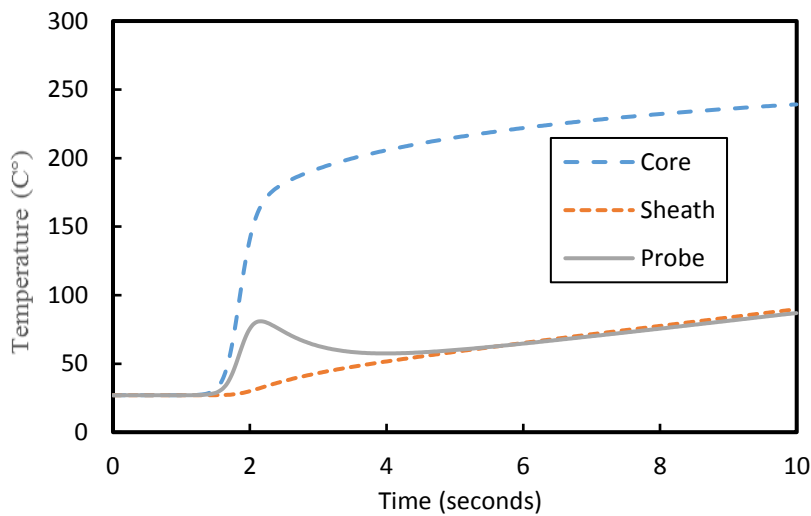

Figure 7: Temperature histories for the probe, sheath and TREAT core during a 2.25$ pulse

Note that the probe temperature converges with the sheath temperature at about the 7 second point also. It would appear that the rise in the sheath temperature is driven by a combination of the probe and the core temperature. The temperature histories for the 1.25$ pulse are given in Figure 8. The rise in the sheath temperature for this case quickly approaches the probe temperature and then drives the probe temperature. Note also in this case that the core temperature continues to increase due to the effects of the delayed neutrons. This can be mitigated by adding negative reactivity to the core after the pulse but that was not of interest to the analysis here.

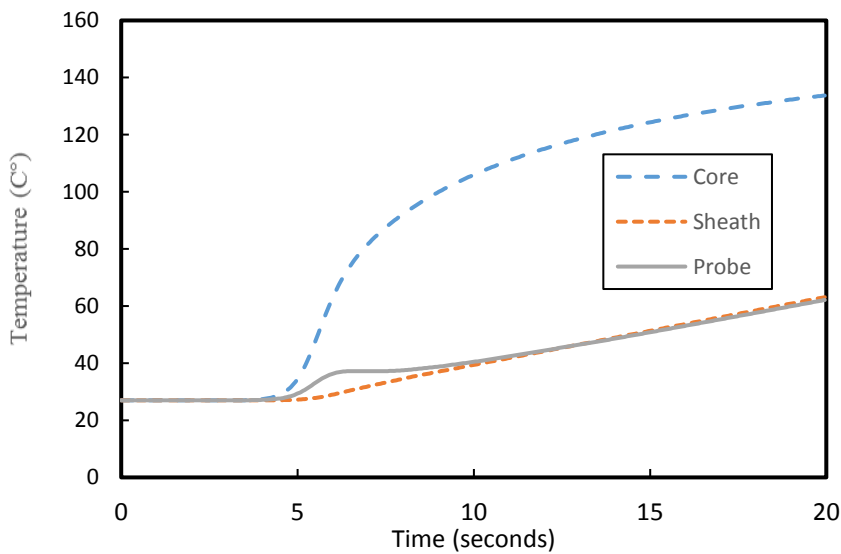

Figure 7: Temperature histories for the probe, sheath and TREAT core during a 1.25$ pulse

This is also a 1-dimensional model, so some analyses will be required to deal with the cosine shape of the flux in the axial direction. It may be possible to combine a series of calculations at points along the length of the probe to get its overall extension, or it may be possible to calibrate the one point it calculates to an finite element analysis (FEA) axisymmetric simulation (Figure 8). The results of the FEA analysis are compared against the available experimental values and the 1-D model in Figure 9.

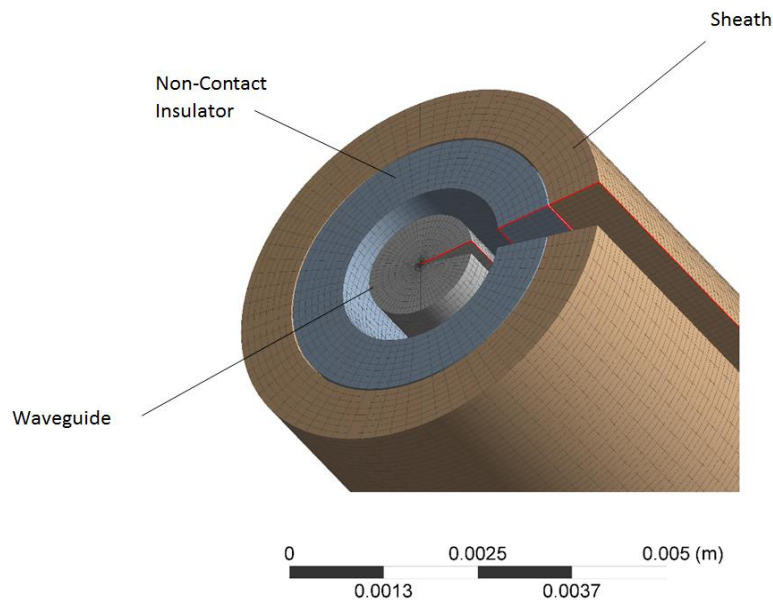

Figure 8: The generated mesh of the probe assembly FEA model

To perform the experimental analysis, an electrically heated experiment on a nichrome (NiCr) wire was designed and built. The nichrome wire was representative of the anticipated uranium-nickel wire that will be used in TREAT. The experiment had 3 different heating rates. The first 480 seconds the wire was heated at the 30 watt level, then the power was upped to 120 watts for 920 seconds, then the power was turned off and the temperatures decayed for 500 seconds. In order to model this experiment the core size in the point kinetics approximation was set to $1.0 \times 10^{15}$ kg and the neutron lifetime increased by 2 orders of magnitude.

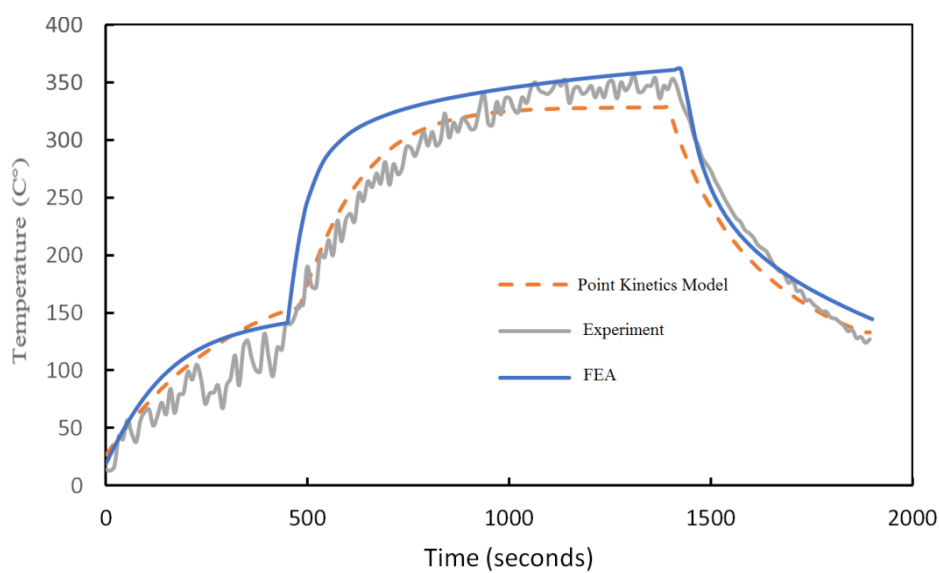

Figure 9: Comparison of the numerical results against the experiment

# 5 CONCLUSION

This article briefly described the transient numerical and experimental analyses performed on the design development of a collaborative project to investigate a novel nuclear heating sensor technology to be applied in the TREAT facility. Results of the experimental probe/sensor heating and computational analyses using two different methods were presented and evaluated. These calculations and analyses were based on the newly designed and fabricated sensor to provide a better understanding of the sensor's behaviour in TREAT. It was observed that the obtained results applied developed point kinetics model, as well as the transient finite element model, are in agreement with the experimental results generated from the electrically heated fabricated sensor. The analyses confirm the benefit of using the point kinetics and FEA in our modelling approach, in support of the sensor design development.